\documentstyle[12pt]{article}
\begin{document}

\begin{center}
{\large {Amplification of External EM-wave by Nonlinear 
Wake Waves in Cold Plasma.}}
\end{center}
\begin{center}
{A.Ts. Amatuni\\
Yerevan Physics Institute, Alikhanian Brothers' str. 2, Yerevan 375036,
Armenia. E-mail: gayane@jerewan1.yerphi.am}
\end{center}
\vspace{0.5cm}

\section{Abstract}

Interaction of external monochromatic, linearly polorized plane EM-wave with 
nonlinear one-dimensional wake wave, generated by relativistic electron bunch
moving in cold plasma, is considered.

At definite conditions on parameters of plasma and bunch, nonlinear wake wave
can have a pronounced spikes, where plasma electron density has its maximum
value and plasma electron velocities are nearing wave breaking limit, i.e. 
velocity of the rigid driving bunch \cite{1}-\cite{5}. Presented calculations 
show, that external EM-wave, propagating through plasma, normal to the bunch 
velocity, at such a conditions can be amplified inside the spikes, due to 
interaction with the plasma electrons. EM-wave, Thomson scattered on spikes,
is also amplified. Amplification factors are obtained for the both cases at 
different conditions on EM-wave-plasma-electron bunch system parameters. An
additional amplification takes place, when the frequency of EM-wave is 
nearing to the spike plasma frequency.

Results are obtained in the perturbative approach, when dimensionless 
external EM-wave amplitude is used as a small parameter. The exact one-
dimensional solution of the problem of nonlinear wake wave generation in 
cold plasma by relativistic electron bunch, obtained previously in \cite{4},
\cite{5}, is taken as a zero order approximation.

\section{Introduction}

The problem of interaction of external electromagnetic wave (EM-wave) with electrons 
or electron bunches, moving in plasma, and with plasma wake waves, generated
by these electron beams, have been a subject of numerous investigations 
\cite{6}-\cite{10}. It was assumed that driving electrons or electron 
bunches are nonrelativistic, generated plasma wake waves are linear and 
physical problems studied in \cite{6}-\cite{10} have been devoted to 
electron ionization losses and electron acceleration.

Plasma - relativistic electron bunch RF - generators have been proposed 
in \cite{12} and experimentally investigated in number of works (see 
reviews \cite{13} - \cite{15}). Nonlinear Thomson scattering of intense 
laser pulses from plasmas was considered in \cite{16}.  

In the present work considered plasma wake waves are essentialy nonlinear,
generated by relativistic rigid electron bunches and external EM-wave is
weak enough in order to provide the possibility to use the perturbative
approach. The goal of present investigation is to find out the possibility 
and conditions, when external EM-wave can be amplified inside the plasma as
well as Thomson scattered EM-wave. Main difference from previous 
investigations consists in consideration the generated plasma waves, 
which are near the wave breaking limit. It is shown that if some conditions 
near
to wake wave breaking limit are satisfied, the essential amplifications in
both considered cases can take place. 

The breaking of waves is one of the
most impressive phenomena, where effect of nonlinearities stands out of 
clearly. Plasma wave breaking was predicted in the works of Akhieser and
Polovin \cite{1} and Dawson \cite{2} (see also \cite{3}). The connection
of the parameters of plasma wake wave breaking limit with the parameters
of plasma and driving relativistic electron bunch in one-dimensional 
approach have been obtained in \cite{4}, \cite{5}. 

Recently, the variety
of nonlinear plasma wave-breaking was considered in the frame of laser
wake field acceleration (LWFA) concept (see e.g. \cite{17}-\cite{19}). Wake 
wave breaking has followed the so called "blow out" regime, \cite{20}
introduced in the frame of plasma wake field acceleration (PWFA) concept
(see also review \cite{19}). 
As it will be seen from the presented results, wake waves near wavebreaking
limit can also be used for EM-wave amplification with possible application 
to desigh the powerful klystron type EM-wave amplifiers or generators for
future linear colliders with high acceleration gradient.

\section{Formulation of the Problem and Basic Equations}

Consider flat electron bunch, moving in cold plasma, along z-direction with 
the velocity $v_0$ in lab system. Longitudinal length of the bunch is d,
transverse dimensions of the bunch are taken infinite, i.e. the problem can
be treated, in the absence of external electromagnetic (EM) field, as one 
dimensional. Bunch is placed at the moment of time $t=0$ in interval 
$0 \le z \le d$, ions are considered as immobile, plasma electron density
at equilibrium is $n_0$, plasma linear frequency $\omega_p^2=4\pi e^2n_0/m$.

One dimensional approaches valid for wide enough bunches, when bunch radius
$r_0 \gg \frac{c}{\omega_p}$ \cite{21}. The condition could be relaced by 
at same sence more adjustable one, if external constant longitudinal strong
enough magnetic field $H_0$ is applied to the system. Then one dimensionality
condition would be $r_0 \gg r_L, \Omega_l \gg \omega_p$, where $r_L=
\frac{v_{tr}}{\Omega_L}=\frac{cp_{tr}}{cH_0}, \Omega_L=\frac{ecH_0}
{\epsilon}$ are Larmour radius and frequency subsequentaly for plasma 
electrons with energy ${\cal E}$, transverse velocity and momenta $v_{tr},
p_{tr}$.

The considered cold plasma - rigid electron bunch system is 
interacting with the external monochromatic, linearly polarized 
electromagnetic wave (EM-wave), propagating through plasma in x direction,
perpendicular to bunch velocity and with electric vector directed along
0z - axis (p - polarization):
\begin{eqnarray}
\label{1}
{\cal E}_z ={\cal E}_{0z}e^{-iw_0t+ik_0x} \\ \nonumber
{{\cal H}}_y = -\sqrt{\epsilon}{\cal E}_{z0}e^{-iw_0t+ik_0x}, k_0=
\sqrt{\epsilon}\frac{\omega_0}{c},
\end{eqnarray}
where $\epsilon$ is a plasma dielectric constant
\begin{equation}
\label{2}
\epsilon = \epsilon'+i\epsilon'' = 1-\frac{\varpi_p^2}{\omega_0^2} + 
i\frac{\varpi_p^2 \nu_{eff}}{\omega_0^3}
\end{equation}
and $\varpi_p^2 = \frac{4\pi e^2n^{(0)}(z)}{m}, n^{(0)}(z)$ is the plasma
electron density in wake wave, generated by driving bunch in the absence of
external EM-wave, $\nu_{eff}$ - effective collisions frequency of plasma
electrons, $\nu_{eff} \ll \varpi_{p} < \omega_0$.

The considered problem consists in finding out the components of EM - field 
with frequency $\omega_0$ inside and outside (Thomson scattered) of the plasma,
which is modification of the external field (\ref{1}) due to interaction with 
the plasma electrons, perturbed by relativistic rigid electron bunch, moving
in cold plasma. The back influence of plasma wake wave on the bunch \cite{22}
as well as bunch interaction with external EM - wave are disregarded, due
to assumption than driving bunch is relativistic enough.

It is assumed also, that considered plasma column has a thickness $\triangle$ 
along the direction of EM - wave propagation (0x - axis), which is smaller
than corresponding skin length of plasma $\triangle < \frac{\sqrt{2}c}
{\omega_0}\sqrt{|\epsilon''|}$.

Introduce the dimensionless arguments:
\begin{equation}
\label{3}
t' = \omega_pt, x'_1z'=k_px, k_pz, k_p=\omega_p/c,
\end{equation}
and dimensionless variables:

\begin{eqnarray}
\label{4}
\vec{E}=\frac{m\omega_pc}{e}\vec{E}', \vec{H} = \frac{m\omega_pc}{e}
\vec{H}',\\ \nonumber
\frac{n_e}{n_0}=n', \frac{n_b}{n_0}=n_b',\\ \nonumber
\beta_0=\frac{v_0}{c}, \vec{\beta}=\frac{\vec{v}_e}{c}, \vec{\rho}=\frac
{\vec{p}_e}{mc}\\ \nonumber
{\cal E}'_z = \frac{m\omega_pc}{e}{\cal E}_z, {\it 
H}'_y=\frac{m\omega_pc} {e}{{\cal H}}_y, \\ \nonumber
{\cal E}'_z = a\frac{\omega_0}{\omega_p}e^{-i\frac{\omega_o}{\omega_p} + 
i\frac{k_0}{k_p}x'}, \\ \nonumber
{{\cal H}}'_y = 
-a\frac{\omega_0}{\omega_p}\sqrt{\epsilon}e^{-i\frac{\omega_0}
{\omega_p}t'+i\frac{k_0}{k_p}x'},\\ \nonumber
a \equiv \frac{e{\cal E}_{0z}}{mc\omega_0} = \frac{e{\cal E}_{z0}}{mc
\omega_p}\frac{\omega_p}{\omega_0},
\end{eqnarray}
where $\vec{E}, \vec{H}$ are EM - field generated in plasma by bunch and plasma 
electrons, $\vec{v}_e, \vec{\rho}_e, n_e$ - velocity, momenta and density of
plasma electrons, a - dimenssionless amplitude of the external EM -wave.

The considered system of cold plasma - rigid relativistic electron bunch - 
external EM - wave is described by the following set of equations (where
primes are temporaly ommited):
\begin{eqnarray}
\label{5}
1.& \frac{\partial{\vec{\rho}}}{\partial{t}}+\left(\vec{\beta}\frac
{\partial}{\partial{\vec{r}}}\right)\vec{\rho}=-\left(\vec{{\cal E}}+
\left[\vec{\beta}{{\cal H}}\right]\right)- \left(\vec{E}+[\beta 
\vec{H}]\right)\\ \nonumber
2.& div \vec{E} = 1-n-n_b \\ \nonumber
3.& \frac{\partial{n}}{\partial t}+ div\left(n\vec{\beta}\right)=0 \\
\nonumber
4.& rot\vec{H}=\frac{\partial \vec{E}}{\partial t}-n\vec{\beta}-n_b
\vec{\beta_0} \\ \nonumber
5.& rot E = -\frac{\partial \vec{H}}{\partial t}
\end{eqnarray}

Continuity eq. (\ref{5}.3) is follows from (\ref{5}.2) and (\ref{5}.4).
For convinience in what follows all equations (\ref{5}) are used.

System (\ref{5}) will be solved by perturbative approach, when 
dimensionless external EM - wave amplitude a is taken as a small parameter,
$a \ll 1$. In the zero approximation $(a=0, {\cal E}=0, {{\cal H}}=0)$ the 
system of equation (\ref{5}) reduces to the set of the following equations,
describing the cold plasma - one dimensional relativistic electron bunch 
system, which were exactly solved in \cite{4},\cite{5}, assuming the 
realization of steady state regime (which means that all variables are 
function of $\tilde{z}=z-\beta_0t$ only):
\begin{eqnarray}
\label{6}
1.& (\beta_0-\beta_z^{(0)})\frac{\partial \rho_z^{(0)}}{\partial 
\tilde{z}}= E_z^{(0)}\\ \nonumber
2.& \frac{\partial E_z^{(0)}}{\partial \tilde{z}} = 1-n^{(0)}(\tilde{z})-n_b
\\ \nonumber
3.& \frac{d}{d\tilde{z}}\left[n^{(0)}\left(\beta_0 - \beta_z^{(0)}
\right)\right]=0
\end{eqnarray}

($\vec{H}=0$ due to the symmetry of the problem).

In (\ref{6}) superscript (0) denote the subsequent variables at zero order 
approximation ($a=0$).

From eq. (\ref{6}.3) with the boundary condition 
$n^{(0)}=n_0, \beta_z^{(0)}=0$, when $\tilde{z}=d$ (on the front of the
moving bunch) it follows that
\begin{eqnarray}
\label{7}
n^{(0)}(\tilde{z})=
\frac{\beta_0}{\beta_0-\beta^{(0)}(\tilde{z})}=
\frac{\beta_0(1+\rho_z^{(0)2})^{1/2}}{\beta_0(1+\rho_z^{(0)2})^{1/2}-
\rho_z^{(0)}} > 0, \\ \nonumber
\beta_z^{(0)}=\frac{\rho_z^{(0)}}{(1+\rho_z^{(0)2})^{1/2}} < \beta_0,
\end{eqnarray}
(this result also can be obtained from eqs. (\ref{5}.2), (\ref{5}.4) written
in zero approximation).

Inside the driving bunch $\beta_z^{(0)}$ is always negative, so $n^{(0)}>0$
automaticaly; behind the bunch, in wake waves, $\beta_z^{(0)}$ periodicaly
changes sign and at some $\tilde{z}$, when $\beta_z^{(0)}$ is $0<\beta_z^{(0)}
\rightarrow \beta_{max} < \beta_0, \tilde{z}\rightarrow \tilde{z}_m$ the 
plasma electron density $n^{(0)}(z_m)$ could be large enough, nearing wave
breaking limit. The values of $\beta_{max}$ (or $\rho_{max}$) could be 
expressed through plasma and driving bunch parameters $n_0,n_b,d, \gamma_0=
(1-\beta_0^2)^{-1/2}$ using results of the work \cite{5}.

For consideration of the next approximations to the set (\ref{5}) introduce 
a new arguments $$\tilde{z}=z-\beta_0t, \tau=t,x=x.$$
Then
\begin{equation}
\label{8}
\frac{\partial}{\partial t}=\frac{\partial}{\partial \tau}- \beta_0\frac
{\partial}{\partial \tilde{z}}, \frac{\partial}{\partial z}= 
\frac{\partial}{\partial \tilde{z}},
\end{equation}

Variables in eqs. (\ref{5}) decompose in the following series
\begin{eqnarray}
\label{9}
\rho_z=&\rho_z^{(0)}(\tilde{z})+\epsilon\rho_z^{(1)}(\tilde{z},x,\tau)+\cdots,
\\ \nonumber
\rho_x=&\epsilon\rho_x^{(1)}(\tilde{z},x,\tau)+\cdots, \\ \nonumber
\beta_z=&\beta_z^{(0)}(\tilde{z})+\epsilon\beta_z^{(1)}(\tilde{z},
x,\tau)+\cdots, \\ \nonumber
\beta_x=&\epsilon\beta_x^{(1)}(\tilde{z},x,\tau)+\cdots, \\ \nonumber
E_z=&E_z^{(0)}(\tilde{z})+\epsilon E_z^{(1)}(\tilde{z}, x, \tau) +\cdots,
\\ \nonumber
E_x=&\epsilon E_x^{(1)}(\tilde{z},x,]tau)+\cdots, \\ \nonumber
H_y=&\epsilon H_y^{(1)}(\tilde{z},x, \tau)+ \cdots,
\end{eqnarray}
where $\epsilon \equiv a$;in what follows $\epsilon$ will be put equal 1 
and a will be included in variables $\vec{\rho}^{(1)}, E^{(1)} ...$.

The variables $\vec{\beta}^{(1)}$ and $\vec{\rho}^{(1)}$ are connected by 
the 
following approximation relation valid, when $|\rho_z^{(1)}| \ll 
|\rho_z^{(0)}|, |\rho_x^{(1)}| \ll |\rho_z^{(0)}|$, up to $O(\epsilon)$
\begin{equation}
\label{10}
\beta_z^{(1)}=\frac{\rho_z^{(1)}}{1+\rho_z^{(0)2})^{3/2}}, \beta_x^{(1)}=
\frac{\rho_x^{(1)}}{(1+\rho_z^{(0)2})^{1/2}}
\end{equation}

In the first approximation $(\sim a)$ the following set of equations is
obtained from (\ref{5}), (\ref{8}), (\ref{9}):
\begin{eqnarray}
\label{11}
1.1.& \frac{\partial \rho_z^{(1)}}{\partial \tau}-(\beta_0-\beta_z^{(0)})
\frac{\partial \rho_z^{(1)}}{\partial \tilde{z}}+\beta_z^{(1)}\frac
{\partial \rho_z^{(0)}}{\partial \tilde{z}}=-E_z^{(1)}-{\cal E}_z \\ 
\nonumber
1.2.& \frac{\partial \rho_x^{(1)}}{\partial \tau}-(\beta_0-\beta_z^{(0)})
\frac{\partial \rho_x^{(1)}}{\partial \tilde{z}}=-E_x^{(1)}+\beta_z^{(0)}
({{\cal H}}_y+H_y^{(1)}) \\ \nonumber
2.& \frac{\partial E_x^{(1)}}{\partial x}+\frac{\partial E_z^{(1)}}
{\partial \tilde{z}} = -n^{(1)} \\ \nonumber
3.& \frac{\partial n^{(1)}}{\partial \tau}-(\beta_0 -\beta_z^{(0)})
\frac{\partial n^{(1)}}{\partial \tilde{z}}+\frac{\partial}{\partial
\tilde{z}}(n^{(0)}\beta_z^{(1)})+\frac{\partial}{\partial x}(n^{(0)}
\beta_x^{(1)})=0 \\ \nonumber
4.1.& \frac{\partial H_y^{(1)}}{\partial x}=-\frac{\partial E_z^{(1)}}
{\partial \tau}-\beta_0\frac{\partial E_z^{(1)}}{\partial \tilde{z}}-
n^{(0)}\beta_z^{(1)}-n^{(1)}\beta_z^{(0)} \\ \nonumber
4.2.& -\frac{\partial H_y^{(1)}}{\partial \tilde{z}}=\frac{\partial 
E_x^{(1)}}
{\partial \tau}-\beta_0\frac{\partial E_x^{(1)}}{\partial \tilde{z}}-n^{(0)}
\beta_x^{(1)} \\ \nonumber
5.& \frac{\partial E_x^{(1)}}{\partial \tilde{z}}-\frac{\partial E_z^{(1)}}
{\partial x}=-\frac{\partial H_y^{(1)}}{\partial \tau}-\beta_0\frac
{\partial H_y^{(1)}}{\partial \tilde{z}}
\end{eqnarray}

System of eqs. (\ref{11}) allowed to search the solution for $\rho_z^{(1)},
\rho_x^{(1)}, E_x^{(1)}, E_z^{(1)}, H_y^{(1)}$ in the forms ($\omega \equiv
\frac{\omega_0}{\omega_p}, k \equiv \frac{k_0}{k_p}$):
\begin{equation}
\label{12}
\rho_z^{(1)}=\rho_{z0}^{(1)}(\tilde{z})e^{-i\frac{\omega_0}{\omega_p}\tau+
i\frac{k_0}{k_p}x}=\rho_{z0}^{(1)}(\tilde{z})e^{-i\omega \tau + ikx}
\end{equation}
(and subsequent expressions for other variables). Obtained set of equations
will contain derivatives on $\tilde{z}$ only, but still is quasilinear and
complicated enough.

In order to simplificate this set of equations further, consider system 
(\ref{11}), (\ref{12}) behind the driving bunch in the region of wake wave
spikes, where $\beta_z^{(0)}$ achived its maximum value $\beta_m > 0$:
\begin{eqnarray}
\label{13}
\beta_z^{(0)} \approx \beta_m + \frac{1}{2}\left(\frac{\partial^2\beta^{(0)}_z}
{\partial \tilde{z}^2}\right)_m(\tilde{z}-\tilde{z}_m)^2, \\ \nonumber
|\tilde{z}-\tilde{z}_m|^2 \ll 2\beta_m/\left(\frac{d^2\beta_t^{(0)}}
{d\tilde{z}^2}\right)_m
\end{eqnarray}

From eqs. (\ref{6}), (\ref{7}) for zero order approximation it follows that
\begin{eqnarray}
\label{14}
\left(\frac{d^2\beta_z^{(0)}}{d\tilde{z}^2}\right)_m=\frac{1}{(1+\rho_m^2)^
{3/2}}\left(\frac{d^2\rho_z^{(0)}}{d\tilde{z}^2}\right)_m, \\ \nonumber
\left(\frac{d^2\rho_z^{(0)}}{d\tilde{z}}\right)_m(\beta_0-\beta_m)=
\left(\frac{\partial E_z^{(0)}}{\partial \tilde{z}}\right)_m=-(n_m-1),
\\ \nonumber
W \equiv \left(\frac{d^2\beta_z^{(0)}}{d\tilde{z}^2}\right)_m=
-\frac{n_m(n_m-1)}{\beta_0(1+\rho_m^2)^{3/2}},
\end{eqnarray}
where $\rho_m, n_m$ are maximun values of plasma electron momenta and 
density at wake wave spikes. If $\beta_m \rightarrow \beta_0, 
\rho_m \gg 1, n_m \gg 1$ and are near to wave breaking limit. The considered 
domain
of the wake wave spikes (\ref{13}), using (\ref{14}), is approximately:
\begin{equation}
\label{15}
|\tilde{z} -\tilde{z}_m| \ll \left(\frac{2\beta_0^2\rho_m^3}{n_m^2}\right)^
{1/2} \approx \frac{2^{1/2}\gamma_0^{3/2}}{n_m}
\end{equation}

In the domain, defined by (\ref{15}), the system of eqs. (\ref{11}), using
(\ref{12}), (\ref{13}) can be essentialy simplified and takes the following
form:
\begin{eqnarray}
\label{16}
1.1.& -i\omega\rho_{z0}^{(1)} + \beta_{z0}^{(1)}\frac{d\rho_z^{(0)}}
{d\tilde{z}}=-E_{z0}^{(1)}-{\cal E}_{z0} \\ \nonumber
1.2.& -i\omega\rho_{x0}^{(1)}=-E_{x0}^{(1)} + \beta_z^{(0)}({{\cal 
H}}_{y0} + H_{y0}^{(1)}) \\ \nonumber
2.& \frac{dE_{z0}^{(1)}}{d\tilde{z}} + ikE_{x0}^{(1)} = -n_0^{(1)} \\ 
\nonumber 
3.& -i\omega n_0^{(1)}+n^{(0)}\frac{d}{d\tilde{z}}\beta_{z0}^{(1)} + 
ikn^{(0)}\beta_{x0}^{(1)} = 0 \\ \nonumber
4.1.& ikH_{y0}^{(1)} = -i\omega E_{z0}^{(1)}-\beta_0\frac{dE_{z0}^{(1)}}
{d\tilde{z}}-n^{(0)}\beta_{z0}^{(1)} - n_0^{(1)}\beta_z^{(0)} \\ \nonumber
4.2.& -\frac{dH_{y0}^{(1)}}{d\tilde{z}} = -i\omega E_{x0}^{(1)} -
\beta_0\frac{dE_{x0}^{(1)}}{d\tilde{z}}-n^{(0)}\beta_{x0}^{(1)} \\
\nonumber
5.& \frac{dE_{x0}^{(1)}}{d\tilde{z}} - ikE_{z0}^{(1)} = i\omega 
H_{y0}^{(1)} + \beta_0 \frac{dH_{y0}^{(1)}}{d\tilde{z}} 
\end{eqnarray}

The system of equations (\ref{16}) is valid up to terms proportional to
$(\tilde{z} - \tilde{z}_m)$, which are small according to assumption 
(\ref{15}), and are disregarded in (\ref{16}).

From eqs. (\ref{16}.4.1) and (\ref{16}.2) it follows
\begin{equation}
\label{17}
H_{y0}^{(1)} = -\frac{\omega}{k}E_{z0}^{(1)} + \beta_z^{(0)}E_{x0}^{(1)} +
\frac{in^{(0)}}{k}\beta_{z0}^{(1)}
\end{equation}

From eqs. (\ref{16}.1.1), (\ref{16}.1.2) and (\ref{17}) it follows:
\begin{equation}
\label{18}
-i\omega \rho_{x0}^{(1)} = \beta_z^{(0)}\left({{\cal H}}_{y0} + 
\frac{\omega}{k}
{\cal E}_{z0}\right) + \frac{\omega}{k}\beta_z^{(0)}\left[-i\omega + 
\frac{in^{(0)}}{\omega}\frac{1}{\left(1+\rho_z^{(0)2}\right)^{1/2}}\right]
\rho_{z0}^{(1)}
\end{equation}

Substituting (\ref{17}), (\ref{16}.4.2) into (\ref{16}.5) it is possible to
obtain
\begin{equation}
\label{19}
E_{x0}^{(1)} = -\frac{k}{\omega}\left(1-\frac{\omega^2}{k^2}\right)
\frac{1}{\left(\beta_0+\beta_z^{(0)}\right)}E_{z0}^{(1)} - 
\frac{i\beta_{z0}^{(1)}n^{(0)}}{k\left(\beta_0 +\beta_z^{(0)}\right)} + 
\frac{i\beta_0n^{(0)}\beta_{x0}^{(1)}}{\omega\left(\beta_0 + 
\beta_z^{(0)}\right)} \end{equation}

In obtaining eqs. (\ref{17} - \ref{19}) the terms proportional to small 
quantities $\beta_0 -\beta_z^{(0)}, 1-\beta_0^2, 1-\beta_z^{(0)2}$ are 
ommited, due to relativism of driving bunch, conditions (\ref{13}) and
$\beta_z^{(0)} \rightarrow \beta_m \rightarrow \beta_0$.

From eqs. (\ref{16}.2), (\ref{16}.1.1), (\ref{18}), (\ref{19}), (\ref{10}) 
it follows:
\begin{eqnarray}
\label{20}
n_0^{(1)} =-\frac{n^{(0)}\rho_{z0}^{(1)}}{\left(\beta_0+\beta_z^{(0)}\right)
\left(1+\rho_z^{(0)2}\right)^{3/2}} + \frac{k}{\omega}\frac{\beta_0n^{(0)}}
{\left(\beta_0+\beta_z^{(0)}\right)\left(1+\rho_z^{(0)2}\right)^{1/2}}
\times \\ \nonumber
\times \left[\frac{i\beta_z^{(0)}}{k}\left({\cal E}_{zo} + \frac{k}{\omega}
{{\cal H}}_{y0}\right)\right. + \\ \nonumber
+\left. \left(\frac{\omega}{k}\beta_z^{(0)} - \frac{\beta_z^{(0)}n^{(0)}}
{\omega k(1+\rho_z^{(0)2})^{3/2}}\right)\rho_{z0}^{(1)}\right] -
\frac{i}{\omega(\beta_0 + \beta_z^{(0)})}(k^2 -\omega^2){\cal E}_{z0} -
\\ \nonumber
-\frac{(k^2 -\omega^2)\rho_{z0}^{(1)}}{\left(\beta_0 + \beta_z^{(0)}\right)}
- i\omega\frac{d\rho_{z0}^{(1)}}{d\tilde{z}} + \rho_{z0}^{(1)}
\left(\frac{d^2\beta_z^{(0)}}{d\tilde{z}^2}\right)_m
\end{eqnarray}

From continuity eq., written in the first approximation (\ref{16}.3), using
(\ref{10}), (\ref{18}) it is possible to obtain:
\begin{eqnarray}
\label{21}
n_0^{(1)} = -\frac{in^{(0)}\beta_z^{(0)}}{\omega(1+\rho_z^{(0)2})^{1/2}}
\left[\left({\cal E}_{z0} + \frac{k}{\omega}{{\cal 
H}}_{y0}\right) 
+ \frac{in^{(0)}}{\omega}\frac{\rho_{z0}^{(1)}}{(1+\rho_z^{(0)2})^{3/2}}-
i\omega\rho_{z0}^{(1)}\right]- \\ \nonumber
-\frac{in^{(0)}}{\omega(1+\rho_z^{(0)2})^{3/2}}\frac{d\rho_{z0}^{(1)}}
{d\tilde{z}}
\end{eqnarray}

From eqs. (\ref{20}) and (\ref{21}) it follows equation for 
$\rho_{z0}^{(1)}$,
where it is possible to take $\beta_z^{(0)}\approx \beta_m \approx \beta_0$
\begin{eqnarray}
\label{22}
-i\omega\left(1-\frac{n^{(0)}}{\omega^2\left(1+\rho_z^{(0)2}\right)^{3/2}}
\right)\frac{d\rho_{z0}^{(1)}}{d\tilde{z}} + \\ \nonumber
+ \left[\frac{n^{(0)2}\beta_0}{2\omega^2\left(1+\rho_z^{(0)2}\right)^2} - 
\frac{n^{(0)}}{2\beta_0\left(1+\rho_z^{(0)2}\right)^{3/2}} - \right. \\ 
\nonumber
 \left. -\frac{n^{(0)}\beta_0}{2\left(1+\rho_z^{(0)2}\right)^{1/2}} +
 \left(\frac{d^2\beta_z^{(0)}}{d\tilde{z}^2}\right)_m - 
\frac{(k^2-\omega^2)} {2\beta_0}\right]\rho_{z0}^{(1)} = \\ \nonumber
= \frac{in^{(0)}\beta_0}{2\omega(1+\rho_z^{(0)2})^{1/2}}\left({\cal E}_{z0}+
\frac{k}{\omega}{{\cal H}}_{y0}\right) +\frac{i(k^2 
-\omega^2)}{2\omega\beta_0} {\cal E}_{z0}
\end{eqnarray}
where ${\cal E}_{z0}, {{\cal H}}_{y0}$ are amplitudes of external EM-wave,
given by (\ref{1}).

\section{Estimates of Amplification Factors}

Coefficients of eq. (\ref{22}) can be estimated using conditions (\ref{13},
\ref{14}, \ref{15}) and in the spikes region defined by (\ref{15}) 
$n^{(0)}\rightarrow n_m \gg 1, \rho_z^{(0)} \rightarrow \rho_m \gg 1$ and 
for $\beta_0 \rightarrow 1$ eq. (\ref{22}) takes simple form
\begin{equation}
\label{23}
A\frac{d\rho_{z0}^{(1)}}{d\tilde{z}}+P\rho_{z0}^{(1)} = Q,
\end{equation}
where
\begin{eqnarray}
\label{24}
A \equiv -i\omega\left(1-\frac{n_m}{\omega^2\rho_m^3}\right) \\ \nonumber
P \equiv \frac{n_m^2}{2\omega^2\rho_m^4}-\frac{n_m}{2\rho_m^3} - \frac{n_m}
{2\rho_m} -\frac{n_m(n_m-1)}{\rho_m^3} -\frac{1}{2}(k^2 - \omega^2) \\
\nonumber
Q \equiv \frac{in_m}{\omega\rho_m}\left({\cal E}_{z0} + \frac{k}{\omega}
{{\cal H}}_{y0}\right) + \frac{i}{2\omega}(k^2 - \omega^2){\cal E}_{z0}
\end{eqnarray}

Assimptotic particular solution of eq. (\ref{23}), which in the spikes 
region (\ref{15}) is independent on $\tilde{z}$ is given by
\begin{equation}
\label{25}
\rho_{z0}^{(1)} \approx Q/P
\end{equation}

The last term in P in (\ref{24}) is 
\begin{equation}
\label{26}
-(k^2 - \omega^2) =\left(\frac{\omega_0}{\omega_p}\right)^2(1-\epsilon)=
\frac{\varpi_p^2}{\omega_p^2}\left(1-\epsilon''\frac{\omega_0^2}{\varpi_p^2}
\right) \approx \frac{\varpi_p^2}{\omega_p}=n_m \gg 1
\end{equation}
and it is much larger than all other terms in P, and when $n_m \gg,
\rho_m \gg 1, \beta_0 - \beta_m \gg \gamma_0^{-3}$
\begin{equation}
\label{27}
P \approx \frac{1}{2}n_m, Q \approx i\left(\frac{n_m}{\rho_m}\frac{\varpi_p^
2}{\omega_0}-\frac{\omega_p}{\omega_0}n_m\right){\cal E}_{z0}
\end{equation}

From (\ref{25})then
\begin{equation}
\label{231}
\rho_{z0}^{(1)} =2i{\cal E}_{z0}\frac{\omega_p}{\omega_0}\left(\frac
{n_m\omega_p}{\rho_m\omega_0}-1/2\right)
\end{equation}

Consider two limiting cases of obtained solution (\ref{231}).

Case (a):
\begin{equation}
\label{241}
\frac{\rho_m\omega_0}{n_m\omega_p} < \frac{\rho_0\omega_0}{n_m\omega_p}=
\frac{(\beta_0-\beta_m)}{(1-\beta_0^2)^{1/2}}\frac{\omega_0}{\omega_p} \ll 1,
\end{equation}
i.e.
\begin{equation}
\label{251}
\beta_0-\beta_m \ll \frac{\omega_p}{\omega_0}\gamma_0^{-1}, n_m = \frac
{\beta_0}{\beta_0-\beta_m} \gg \frac{\omega_0}{\omega_p}\gamma_0
\end{equation}

Simultaneous fullfilment of the condition of plasma spikes transparency 
$\frac{\varpi_p^2}{\omega_0^2} < 1, n_m < \frac{\omega_0^2}{\omega_p^2}$
could take place if:
\begin{eqnarray}
\label{261}
1 \ll \frac{\omega_0}{\omega_p}\gamma \ll n_m < \frac{\omega_0^2}{\omega_p^2},
1 \ll \gamma_0 \ll \frac{\omega_0}{\omega_p}, \\ \nonumber
\frac{\omega_p^2}{\omega_0^2} < \beta_0-\beta_m \ll \frac{\omega_p}{\omega_0}
\gamma_0^{-1};
\end{eqnarray}

Case (b):
\begin{equation}
\label{271}
\frac{n_m\omega_p}{\rho_m\omega_0} \ll 1
\end{equation}

It is opposite to the case (a) condition and it takes place, when
\begin{eqnarray}
\label{28}
\beta_0-\beta_m \gg \frac{\omega_p}{\omega_0}\gamma_0^{-1}, 
\beta_0-\beta_m > \frac{\omega_p^2}{\omega_0^2} \\ \nonumber
1 \ll n_m \ll \frac{\omega_0}{\omega_p}\gamma_0, 1\ll n_m \le 
\frac{\omega_0^2}{\omega_p^2}
\end{eqnarray}

In the case (b) no additional restriction on $\gamma_0$ is needed, the 
relativistic condition $ \gamma_0 \gg 1$ remains valid.

Wake spikes domain, given by (\ref{15}), when $n_m$ is taken by order of 
magnitude equal $\omega_0^2/\omega_p^2$ is
\begin{equation}
\label{29}
|\tilde{z}-\tilde{z}_m| \ll 2^{1/2}\gamma_0^{3/2}\frac{\omega_p^2}
{\omega_0^2}
\end{equation}
and in ordinary units is much smaller than $\lambda_p/2\pi$ in the
case (a) and in the case (b) could be by order of magnitude equal to 
$\lambda_p/2\pi$ , where $\lambda_p$ is plasma linear wave length.

Electric field z - component inside the plasma wake wave spikes (\ref{15}),
using (\ref{16}.1.1), (\ref{231}) is
\begin{equation}
\label{30}
E_{z0}^{(1)} \approx -{\cal E}_{z0} - \frac{i\omega_0}{\omega_p}\rho_{z0}^
{(1)} = -2{\cal E}_{z0}\left(1-\frac{n_m\omega_p}{\rho_m\omega_0}\right)
\end{equation}
and for the case (a) is given by 
\begin{equation}
\label{31}
E_{z0}^{(1)} \approx 2{\cal E}_{z0}\left(\frac{n_m\omega_p}{\rho_m\omega_0}
\right) \gg {\cal E}_{z0}
\end{equation}

For the case (b):
\begin{equation}
\label{32}
E_{z0}^{(1)} \approx -2 {\cal E}_{z0}
\end{equation}
and field amplification factors are
\begin{equation}
\label{33}
K_z=\frac{|E_{z0}^{(1)}|}{|{\cal E}_{z0}|}, K_z^a=2\left(\frac{n_m\omega_p}
{\rho_m\omega_0}\right) \gg 1, K_z^b = 2
\end{equation}
for the case (a) and case (b) correspondingly.

In order to obtain x - component of the electric field inside the spikes 
it is necessary to use eqs. (\ref{18}), (\ref{19}). Assuming that ratio
$\frac{n_m\omega_p}{\rho_m^3\omega_0}$ is small for $\rho_m \gg 1$ and using
(\ref{231}) it is possible to obtain
\begin{eqnarray}
\label{34}
\rho_{x0}^{(1)} \approx \frac{\i\rho_0}{k}\left({\cal E}_{z0}+\frac{k}
{\omega}{{\cal H}}_{z0}\right) +\beta_0{\omega}/{k}\rho_{z0}^{(1)} 
\approx \\ \nonumber
\approx \frac{i}{\sqrt{\epsilon}}\frac{\omega_p}{\omega_0}(1-\epsilon'){\cal 
E}_{z0} + \frac{1}{\sqrt{\epsilon}}\rho_{z0}^{(1)} = \\ \nonumber
=\frac{i}{\sqrt{\epsilon}}
{\cal E}_{z0}\left\{n_m\left(\frac{\omega_p}{\omega_0}\right)^3 + \frac
{2\omega_p}{\omega_0}\left(\frac{n_m\omega_p}{\rho_m\omega_0}-1/2\right)
\right\} 
\end{eqnarray}

From eq (\ref{19}) at the same condition it follows
\begin{equation}
\label{35}
E_{xo}^{(1)}=-\frac{1}{2\beta_0k\omega}(k^2-\omega^2)E_{z0}^{(1)}+ \frac
{i}{\omega}\frac{n^{(0)}\rho_{x0}^{(1)}}{2(1+\rho_z^{(0)2})^{1/2}}
\end{equation}
and using (\ref{30}), (\ref{34}) after some transformation it is possible to
obtain
\begin{equation}
\label{36}
E_{x0}^{(1)}=-\frac{{\cal E}_{z0}}{2\sqrt{\epsilon}}\frac{\omega_p}{\omega_0}
\left[2n_m\left(\frac{\omega_p}{\omega_0}\right)\left(1-\frac{n_m\omega_p}
{\rho_m\omega_0}\right) - \frac{n_m\omega_p}{\rho_m\omega_p}\left(n_m\frac
{\omega_p^2}{\omega_0^2}+ 2\frac{n_m\omega_p}{\rho_m\omega_0}-1\right)\right]
\end{equation}

For the case (a), when $\frac{n_m\omega_p}{\rho_m\omega_0} \gg 1, \frac
{\omega_p}{\omega_0} \gg 1$ and $n_m \le \frac{\omega_0}{\omega_p}$
\begin{equation}
\label{37}
E_{x0}^{(1)} \approx \frac{{\cal E}_{z0}}{\sqrt{\epsilon}}\frac{n_m^2}
{\rho_m}\left(\frac{\omega_p}{\omega_0}\right)^3=\frac{{\cal E}_{z0}}
{\sqrt{\epsilon}}\left(\frac{n_m\omega_p}{\rho_m\omega_0}\right)n_m
\left(\frac{\omega_p}{\omega_0}\right)^2 \le \frac{{\cal E}_{z0}}
{\sqrt{\epsilon}}\left(\frac{n_m\omega_p}{\rho_m\omega_0}\right)
\end{equation}
and amplification factor
\begin{equation}
\label{38}
K_x^a=\frac{|E_{x0}^{(1)}|}{|{\cal E}_{z0}|} = \frac{1}{\sqrt{\epsilon}}
\frac{n_m^2}{\rho_m}\left(\frac{\omega_p}{\omega_0}\right)^3
\end{equation}
could be large enough, if
\begin{equation}
\label{39}
\left(\frac{\omega_p}{\omega_0}\right)^2<(\beta_0-\beta_m) \ll 
\min\left\{\frac{1}{\gamma^{1/2}}\left(\frac{\omega_p}{\omega_0}\right)^{3/2},
\frac{1}{\gamma}\left(\frac{\omega_p}{\omega_0}\right)\right\}
\end{equation}

Conditions (\ref{39}) included conditions (\ref{251}) for the realization of 
the case (a). For the case (b), when $\left(\frac{n_m\omega_p}
{\rho_m\omega_0}\right) \ll 1$ and $1 \ll n_m\frac{\omega_p}{\omega_0} 
\ll \rho_m$:
\begin{equation}
\label{40}
E_{x0}^{(1)}=-\frac{{\cal E}_{z0}}{\sqrt{\epsilon}}n_m\left(\frac{\omega_p}
{\omega_0}\right)^2
\end{equation}
and corresponding field amplification factor is 
\begin{equation}
\label{41}
K_x^b=\frac{1}{\sqrt{\epsilon}}n_m\left(\frac{\omega_p}{\omega_0}\right)^2,
n_m \le \left(\frac{\omega_0}{\omega_p}\right)^2
\end{equation}

In all considered cases it is necessary to be sure that adopted perturbative
approach is not violated. In particular, the conditions for applicability of
the decompositions (\ref{9}), (\ref{10}) must be fulfilled
\begin{equation}
\label{42}
|\rho_{z0}^{(1)}| \ll |\rho_z^{(0)}|, |\rho_{x0}^{(1)}| \ll |\rho_z^{(0)}
\end{equation}

In the considered plasma wake wave spikes domain (\ref{15}) the conditions 
(\ref{42}) using (\ref{4}), (\ref{231}) can be rewritten as
\begin{equation}
\label{43}
\rho_m \gg \left|2a\left(\frac{n_m\omega_p}{\rho_m\omega_0}-1/2\right)
\right|
\end{equation}

For the case (a) it means that:
\begin{eqnarray}
\label{44}
1 \ll \frac{n_m\omega_p}{\rho_m\omega_0} \ll \frac{\rho_m}{2a}, 
a \equiv \frac{e{\cal E}_0}{mc\omega_0} \ll 1 \\ \nonumber
a \ll \frac{\rho_m^2\omega_0}{2n_m\omega_p}
\end{eqnarray}
For the case (b) condition (\ref{43}) just gives
\begin{equation}
\label{45}
a \ll \rho_m, \rho_m \gg 1
\end{equation}

Turning now to estimates of amplification factors for EM-wave, Thomson 
scattered on plasma spikes, consider the plasma current densities on spikes,
which in the first approximation of considered perturbative approach are 
given by
\begin{equation}
\label{46}
j_{x0}^{(1)} = n^{(0)}\beta_{x0}^{(1)}, j_{z0}^{(1)}=n^{(0)}\beta_{z0}^{(1)}
+n_0^{(1)}\beta_z^{(0)}
\end{equation}

In considered plasma wake wave spikes domain (\ref{15}):
\begin{equation}
\label{47}
n^{(0)} \approx n_m, \beta_z^{(0)} \approx \beta_m \approx \beta_0 \approx 1,
\beta_{z0}^{(1)} \approx \frac{\rho_{z0}^{(1)}}{\rho_m^3}, \beta_{x0}^{(1)}
\approx \frac{\rho_{x0}^{(1)}}{\rho_m}
\end{equation}

From eq. (\ref{21}) for $n_0^{(1)}$ it follows that the main contribution to
$n_0^{(1)}$ came in the case (a) from the last term in the square bracket
in (\ref{21}) and, using (\ref{231}) for the case (a), obtain:
\begin{equation}
\label{48}
n_0^{(1)} \approx -\frac{n^{(0)}\beta_z^{(0)}}{\left(1+\rho_z^{(0)2}\right)^
{1/2}}\rho_{z0}^{(1)} \approx -2i\left(\frac{n_m\omega_p}{\rho_m\omega_0}
\right)^2{\cal E}_{z0}
\end{equation}

For considered current densities (\ref{46}) using (\ref{47}, \ref{48}, 
\ref{231}, \ref{34}) the following estimates for the case (a) can be 
obtained: \begin{eqnarray}
\label{49}
j_{x0}^{(1)} \approx n_m\frac{\rho_{x0}}{\rho_m} \approx 
\frac{2}{\sqrt{\epsilon}}
\left(\frac{n_m\omega_p}{\rho_m\omega_0}\right)^2{\cal E}_{z0} \\ \nonumber
j_{z0}^{(1)} \approx n_0^{(1)} \approx -2i\left(\frac{n_m\omega_p}
{\rho_m\omega_0}\right)^2{\cal E}_{z0}
\end{eqnarray}

The radiated outside the plasma Thomson scattered EM-wave is described by
potentials:
\begin{equation}
\label{50}
A_{x,y}=\frac{1}{cR_0}\int{j_{x,y}(x,y,z,t-\frac{\vec{r}\vec{n}}{c})}dv \sim
\frac{j_{0x,y}^{(1)}}{cR_0},
\end{equation}
where $R_0$ is the distance from radiator (spikes) to observation point
outside the plasma, $r^2=x^2+y^2+z^2, \vec{n}$ - unit vector in direction of
radiation. Radiated energy flux in the unit solid angle per second from unit
volume of plasma spike is proportional to 
$$\frac{dW_{x,z}}{d\Omega} \sim |j_{0x,z}^{(1)}|^2$$
and corresponding intensity amplification factor
\begin{equation}
\label{51}
K_{rad x,z}=\frac{1}{W_0}\frac{dW_{x,z}}{d\Omega} \sim \frac{|j_{0x,z}|^2}
{|{\cal E}_{z0}|^2}
\end{equation}

In ({\ref{51}) $W_0$ is the incident energy flux of external EM - wave 
(\ref{1})
on unit area of plasma wake wave spike cross section normal to 0x -axis 
(direction of EM - wave propagation) per second, $W_0 \sim |{\cal E}_{z0}|^2$.
From (\ref{51}), (\ref{49}) it follows that
\begin{equation}
\label{52}
K_{rad x}^a \sim \frac{4}{|\epsilon|}\left(\frac{n_m\omega_p}{\rho_m\omega_0}
\right)^4 \gg 1
\end{equation}
\begin{equation}
\label{53}
K_{rad z}^a \sim 4 \left(\frac{n_m\omega_p}{\rho_m\omega_0}\right)^4 \gg 1
\end{equation}

The field amplification factors (\ref{38}), (\ref{41}) and intensity 
amplification factor (\ref{53}), attributed to the plasma electron motion in
plasma wave spikes along x - axis posses multipliers $|\epsilon|^{-1/2}$ 
and 
$|\epsilon|^{-1}$ subsequently. It means that at resonance conditions, when
frequency of external
EM - wave is nearing the plasma frequency at the nonlinear wave spikes 
$\omega_0^2 \rightarrow \varpi_p^2 = n_m\omega_p^2$ an additional 
amplification can take place. Additional amplification factors are equal,
according to (\ref{2}):
\begin{eqnarray}
\label{54}
|\epsilon|^{-1/2}=\frac{1}{\sqrt{|\epsilon''|}}=\left(\frac{\omega_0}
{\nu_{eff}}\right)^{1/2} \\ \nonumber
|\epsilon|^{-1}=\frac{1}{\sqrt{|\epsilon''|}}=\left(\frac{\omega_0}
{\nu_{eff}}\right), \omega_0 \rightarrow \varpi_p, \nu_{eff} \ll \omega_p
\ll \omega_0
\end{eqnarray}

Mentioned resonance amplification (\ref{54}) takes place, of course 
independantly from conditions (\ref{251}), (\ref{28}) of realization the 
cases (a) or (b).

\section{Conclusion}

The obtained analytical estimates demonstrate, that at certain conditions
on cold plasma - relativistic electron bunch - external EM - wave system 
parameters (see (\ref{251}), (\ref{28}), (\ref{54})), essentialy large
amplification of electric field inside the plasma spikes, as well as 
intensity of Thomson scattered on spikes EM - wave, are existed (see
expression for amplification factors (\ref{33}), (\ref{38}), (\ref{41}),
(\ref{52}), (\ref{53}), (\ref{54})). Presented results could be used in 
research and 
development of powerful klystron type amplifiers and generators of high
frequency EM - waves for future linear colliders.

The presented in the work estimates have at some extent qualitative character
and must be complemented by more quantative investigations, presumably by
computer simulations. The reason for computer calculations is evident from the
fact, that even in perturbative approach, addopted in present work, the 
problem is reduced, due to nonlinearity of zero order approximation, to the 
set of quasilinear equations (\ref{11}), with variable coeficients, nonlinearly
depending on argument $\tilde{z}$. The exact analytical solution of the set of
equations (\ref{11}) is practicaly impossible to obtain.

However, more elaborate perturbative approach, for example based on multiple
scales method, could provide the possibility to go further in directions
outlined in the present work.

In order to find out the optimal geometry for experimental detection of the 
predicted EM - wave amplification, it is necessary to consider also different
directions of propagation and different polarizations of external EM - wave.

\end{document}